\documentclass[review,12pt,authoryear]{elsarticle}

\usepackage{amssymb}
\usepackage{lineno}
\usepackage{titlesec}
\usepackage{natbib}
\setcitestyle{numbers,square}
\titleformat{\subsection}[block]{\normalsize\bfseries}{\arabic{section}.\arabic{subsection}}{1em}{}[]
\usepackage[colorlinks,linkcolor=blue,anchorcolor=blue,citecolor=blue]{hyperref}
\setcitestyle{square,citesep={,},aysep={},yysep={;}}
\usepackage{fancyhdr}

\journal{ }

\begin{document}
% \linenumbers

\begin{frontmatter}

% \maketitle

\title{Signal significance incorporating systematic uncertainty for continuous test}


\author[author1]{Yi Ding}
\author[author1]{Weiming Song}
\affiliation[author1]{Jilin University, Changchun Jilin, China}

\author[author2]{Kai Zhu}
\affiliation[author2]{Institute of High Energy Physics, Beijing, China}

\date{ }

\begin{abstract}
To properly estimate signal significance while accounting for both statistical and systematic uncertainties, we conducted a study to analyze the impact of typical systematic uncertainties, such as background shape, signal shape, and the number of backgrounds, on significance calculation using the continuous test method. Our investigation reveals unexpected and complex features, leading us to recommend a conservative approach: one should estimate signal significance by conducting trials with as many as possible combinations of various uncertainties associated with the fitting procedure, and then select the ``worst" outcome as the final result.
\end{abstract}
% \maketitle
% \oddfoot{\footnotesize\itshape
%       Preprint submitted to \ifx\@journal\@empty Elsevier
%       \else\@journal\fi\hfill\today}%

% \begin{keyword}



% \end{keyword}

\end{frontmatter}






\section{Introduction}
Scientists often lay claim to new discoveries, yet quantifying the extent of their deviation from established knowledge remains a persistent challenge. In last decades, the concept of ``significance" has gained widespread adoption in scientific disciplines, particularly in physics, economics, biology, psychology, sociology, and medicine. It serves as a means to gauge the confidence with which a scientist asserts a new discovery. For instance, in high-energy physics, a conventional ``rule" for significance is based on standard deviations ($\sigma$), where a signal with a significance of $3\sigma$ is considered as evidence, and only one with a significance of $5\sigma$ is deemed an observation of note. The paper~\cite{Sinervo:2002sa} by P.~K.~Sinervo provides a historical review of the concept of ``statistical significance" in observations and its application in high-energy experiments. Accurate estimation of significance is crucial in data analysis, as overestimation can damage one's reputation, while underestimation may lead to overlooking important discoveries. A ``significant" observation typically allows for the elimination of one or more hypotheses in favor of alternatives within a statistical framework. Hence, in essence, this constitutes a statistical problem, which is why significance is sometimes referred to as ``statistical significance". Two fundamental methods complement the calculation of significance: the counting method and the continuous test method. This paper follows the definition and calculation of statistical significance proposed in Ref.~\cite{zhusy}, which establishes a correlation between the normal distribution integral probability and the observed p-value. This approach yields explicit expressions for both counting experiments and continuous test statistics. For counting experiments,
\begin{equation}
\label{eq:1}
\int^S_{-S} N(0,1) = 1-p(n_{obs}) = \sum^{n_{obs}-1}_{n=0} \frac{b^n}{n!}e^{-b}\ ,
\end{equation}
where $S$, $n_{obs}$, $b$, and $N(0,1)$ are the signal significance, number of observed events, number of backgrounds, and normal distribution, respectively. For continuous test,
\begin{equation}
\label{eq:2}
\int^S_{-S} N(0,1) = 1-p(t_{obs}) = \int^{t_{obs}}_0 \chi^2(t;r)dt \ ,
\end{equation}
where $t_{obs}\equiv 2[\mathrm{ln} L_{max}(s+b) - \mathrm{ln} L_{max}(b)] $ with $L$ is the likelihood obtained from fits, $L_{max}(s+b)$ is the maximum likelihood value with both signal and background in the fit and $L_{max}(b)$ is the maximum likelihood value with background in the fit. The $r$ is the difference in numbers of freedom.

The significance calculation may initially appear to be a mere statistical exercise, involving the numbers of observed events, backgrounds, and signals, as well as the uncertainties associated with these figures. However, it became evident that this calculation must extend beyond pure statistical analysis, as there are always systematic uncertainties related to factors such as the number of backgrounds, efficiency, signal shape, and background shape. These additional uncertainties introduce new complexities and difficulties to the significance calculation. In the case of the counting method, some proposed discussions on how to account for systematic uncertainty in significance calculation have been put forward~\cite{Bityukov:2002eq,Cousins:2007yta}. The central idea in these papers involves varying the conditions during the calculation based on the systematic uncertainty, and then selecting the ``worst" (i.e., the least) significance as the final result. However, to the best of our knowledge, the consideration of systematic uncertainty in the continuous test method remains unaddressed. Therefore, this paper aims to explore the impact of systematic uncertainties on signal significance under various typical conditions and seeks to provide a solution for properly estimating signal significance when employing the continuous test method in the presence of systematic uncertainty.

\section{Effects to significance due to systematic uncertainties}
We determine the signal significance using a continuous test approach with a toy model constructed from ad hoc invariant mass distributions. In this model, the data consists of two components, namely signal and background, and is generated with specific numbers and distributions. We assess the impact of three types of systematic uncertainties related to the background shape, number of backgrounds, and signal shape.

\begin{figure}[htbp]
\begin{center}
\begin{minipage}[t]{0.6\linewidth}
\includegraphics[width=1\textwidth]{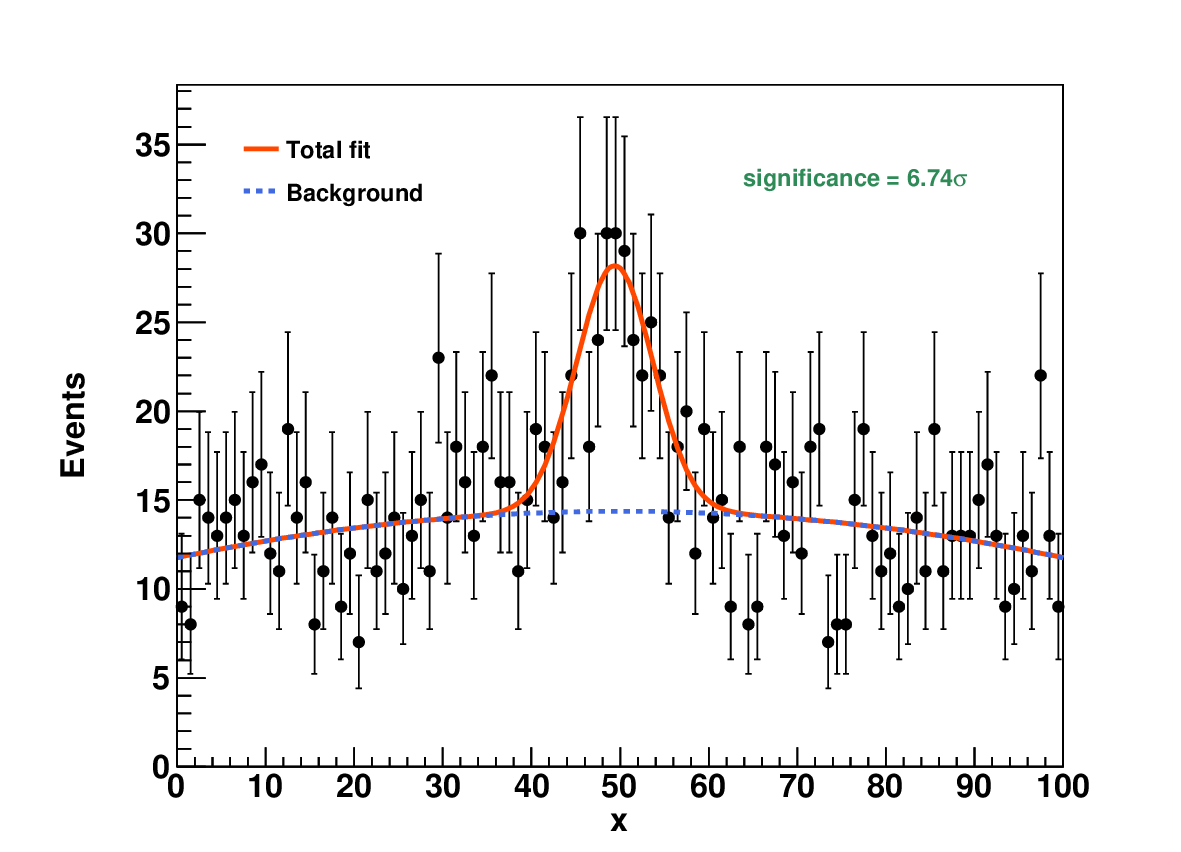}
\end{minipage}%
\caption{The fit of sample by generating function.}
\label{fig:nomalfit}
\end{center}
\end{figure}

To begin, we generate a one-variable sample comprising 1500 events using the following formula
\begin{equation}
\label{eq:3}
    F(x) = (1-f)\cdot G(m,\sigma;x) + f\cdot P(a_0, a_1;x) \ ,
\end{equation}
where $G(m,\sigma;x)$ and $P(a_0, a_1;x)$ are Gaussian and Polynomial functions, and $f$ is the ratio of the background. The values of the corresponding parameters are listed in Table~\ref{tab:data}. This sample will be used in the following for further studies. A fit to the generated sample, with $F(x)$ described by Eq.~\ref{eq:3}, is presented in Fig~\ref{fig:nomalfit}. And the statistical significance is estimated to be $6.7~\sigma$  by the continuous test method that is described by Eq.~\ref{eq:2}. 

\begin{table}[hbtp]
    \centering
    \begin{tabular}{c|c|c|c|c|c}
    $N_{evt}$ & $f$ & $m$ & $\sigma$ & $a_0$ & $a_1$   \\ \hline
    1500    &  $0.9$& $50$ & $5$   &  $0.0$ & $-0.1$ \\  
    \end{tabular}
    \caption{Parameters used to generate the toy MC sample.}
    \label{tab:data}
\end{table}

\subsection{Background shape}
\begin{figure}[htbp]
\begin{center}
\begin{minipage}[t]{0.6\linewidth}
\includegraphics[width=1\textwidth]{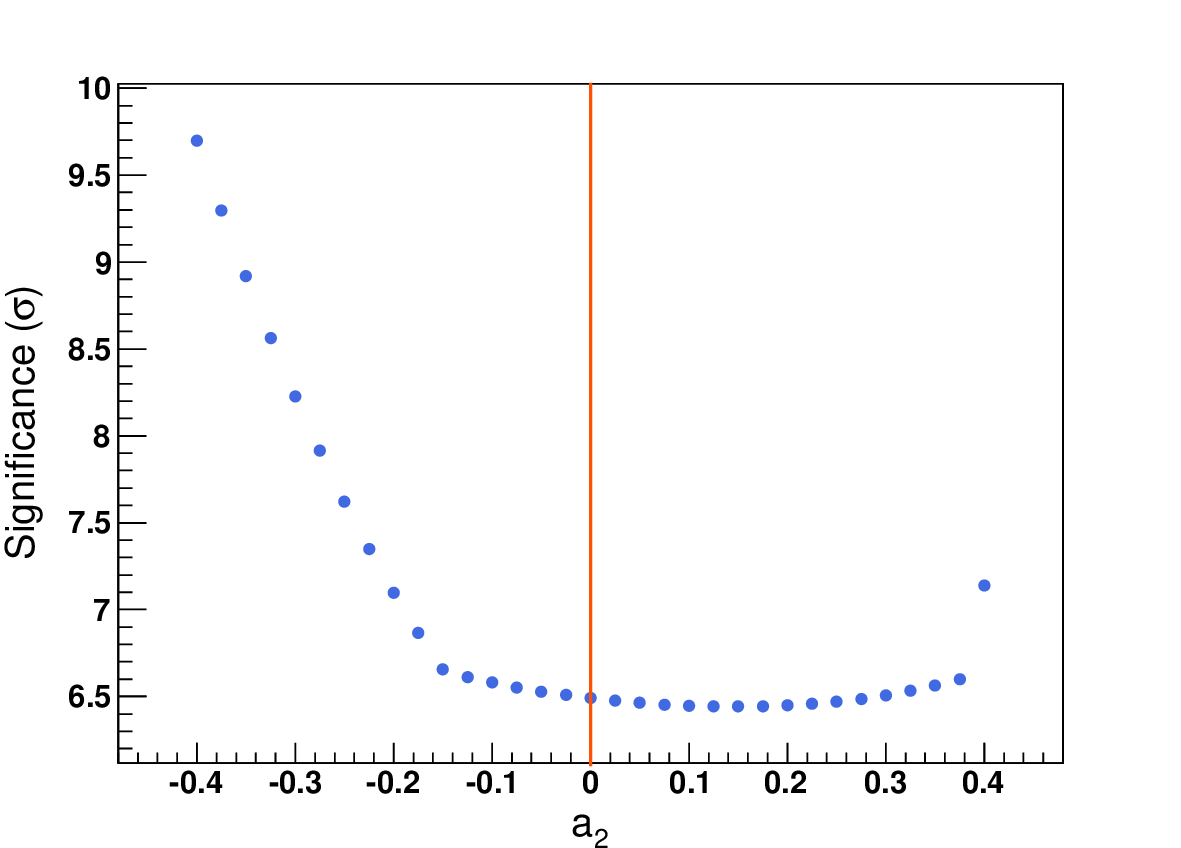}
\end{minipage}%
\caption{The significance distribution is plotted with respect to different background shapes, which are modified by an additional polynomial term indicated by coefficient $a_2$. The vertical line represents the position where the fit function is reduced to the generating function, i.e., when $a_2=0$.}
\label{fig:bkgshape}
\end{center}
\end{figure}

We investigate the impact of systematic uncertainty associated with the background shape by fitting the generated sample using the formula:
\begin{equation}
    F(x) = (1-f)\cdot G(m,\sigma;x) + f\cdot P(a_0, a_1, a_2;x) \ .
\end{equation}
This formula is identical to the one used to generate the sample, except for an additional higher-order term in the polynomial series represented by its coefficient $a_2$. We determine the significance of the signal by applying Eq.~\ref{eq:2}, with $a_2$ varied from $-0.4$ to $0.4$ in increments of $0.025$. It is important to note that the fit function reduces to the generating function when $a_2=0$. The resulting significance is presented in Fig.~\ref{fig:bkgshape} as a function of $a_2$.

From Fig.~\ref{fig:bkgshape}, it is evident that the background shape can have an unpredictable impact on the signal significance in both direction and magnitude. Within the parameter space, the significance changes gradually around $a_2=0$, but increases rapidly when the value of $a_2$ significantly deviates from the true value, regardless of whether it is larger or smaller than the true value. In such cases, the fit function is unable to adequately describe the data. In summary, the significance may either increase or decrease, and establishing a correlation between changes in background shape and changes in significance appears to be extremely difficult.

\subsection{Number of background events} 
\begin{figure}[htbp]
\begin{center}
\begin{minipage}[t]{0.6\linewidth}
\includegraphics[width=1\textwidth]{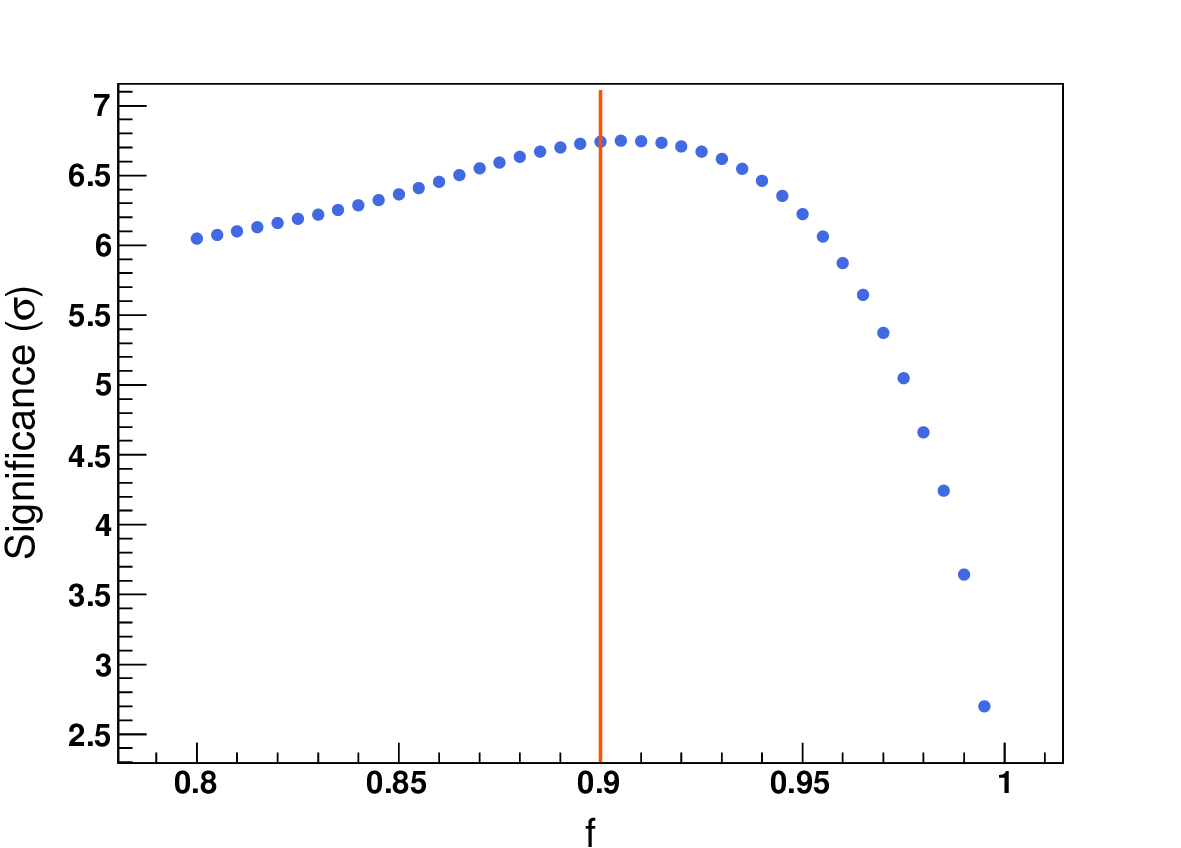}
\end{minipage}%
\caption{The significance distribution is plotted with respect to different background configurations, which are set with a sequence of varied values. The vertical line represents the position where the fit function matches the generating function, i.e., when $f=0.9$.}
\label{fig:bkgcontri}
\end{center}
\end{figure}

We investigate the impact of systematic uncertainty associated with the number of background events by fitting the generated sample the the generating function, while the background ratio $f$ is varied from $0.8$ to $1.0$ in increments of $0.005$.  It is important to note that the fit function reduces to the generating function when $f=0.9$. The resulting significance is presented in Fig.~\ref{fig:bkgcontri} as a function of $f$.

From Fig.~\ref{fig:bkgcontri}, we observe that the number of background events can unexpectedly impact the signal significance. For instance, when the proportion of background events increases slightly, the significance also increases, contrary to our expectations of a decrease. This feature makes it nearly impossible to predict the variation in significance in relation to changes in the number of background events.

\subsection{Signal shape} 
\begin{figure}[htbp]
\begin{center}
\begin{minipage}[t]{0.6\linewidth}
\includegraphics[width=1\textwidth]{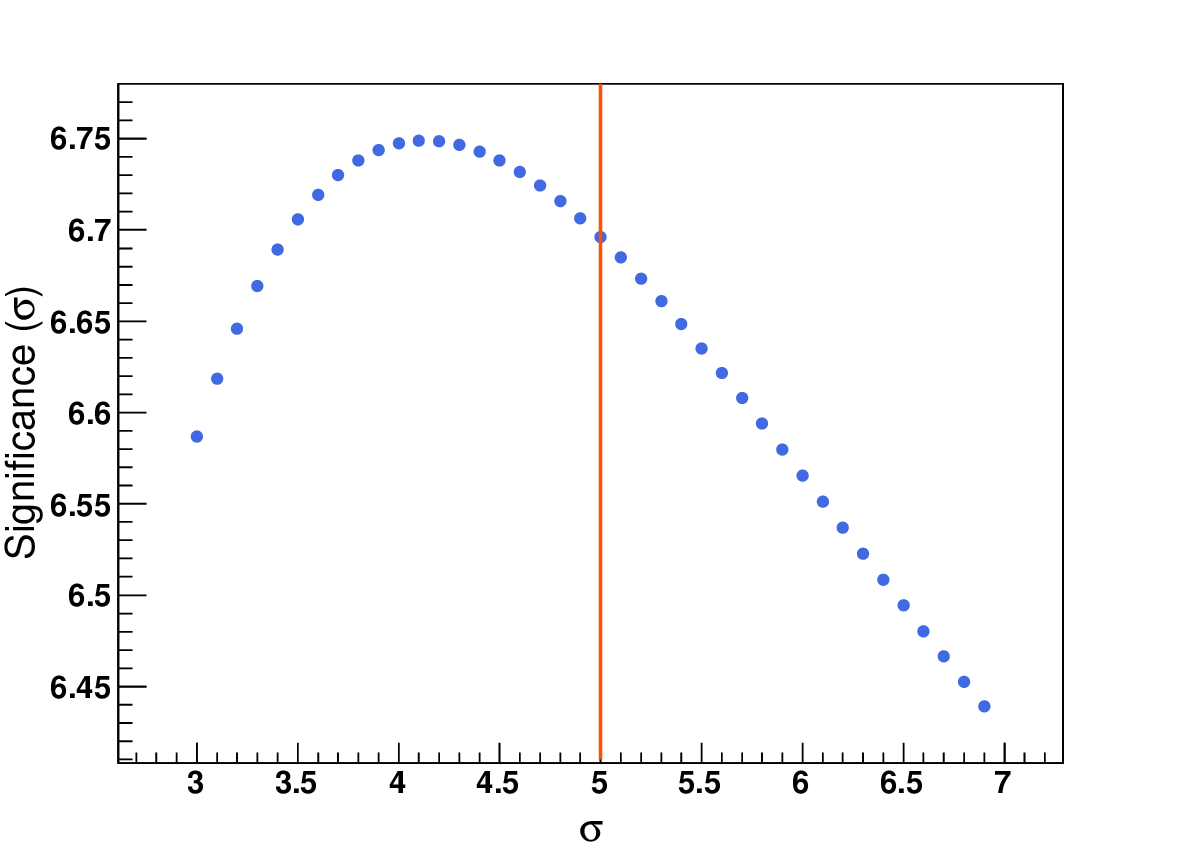}
\end{minipage}%
\caption{The significance distribution is plotted with respect to different signal width, which are set with a sequence of varied values. The vertical line represents the position where the fit function matches the generating function, i.e., when $\sigma=5.0$.}
\label{fig:sigshape}
\end{center}
\end{figure}

We investigate the impact of systematic uncertainty associated with the signal shape by fitting the generated sample the the generating function, while the $\sigma$, that describe the width of the signal, is varied from $3.0$ to $7.0$ in increments of $0.1$. It is important to note that the fit function reduces to the generating function when $\sigma=5.0$. The resulting significance is presented in Fig.~\ref{fig:sigshape} as a function of $\sigma$.

From Fig.~\ref{fig:sigshape}, it is evident that in the vicinity of the true value, the significance increases as the signal shape decreases, while it exhibits the opposite trend when the signal shape increases. This aligns with our expectations, as a narrower signal width facilitates easier discrimination between signal and background. However, when the signal width becomes very narrow, the significance decreases. We attribute this behavior to the fit function's inability to accurately capture the generated sample under such circumstances. Despite the seemingly linear correlation between the signal width and the significance, we conclude that it would be overly ambitious to make any qualitative predictions about the significance based on variations in the signal shape.

\section{Comparison between counting method and continuous test method}
\begin{figure}[htbp]
\begin{center}
\begin{minipage}[t]{0.6\linewidth}
\includegraphics[width=1\textwidth]{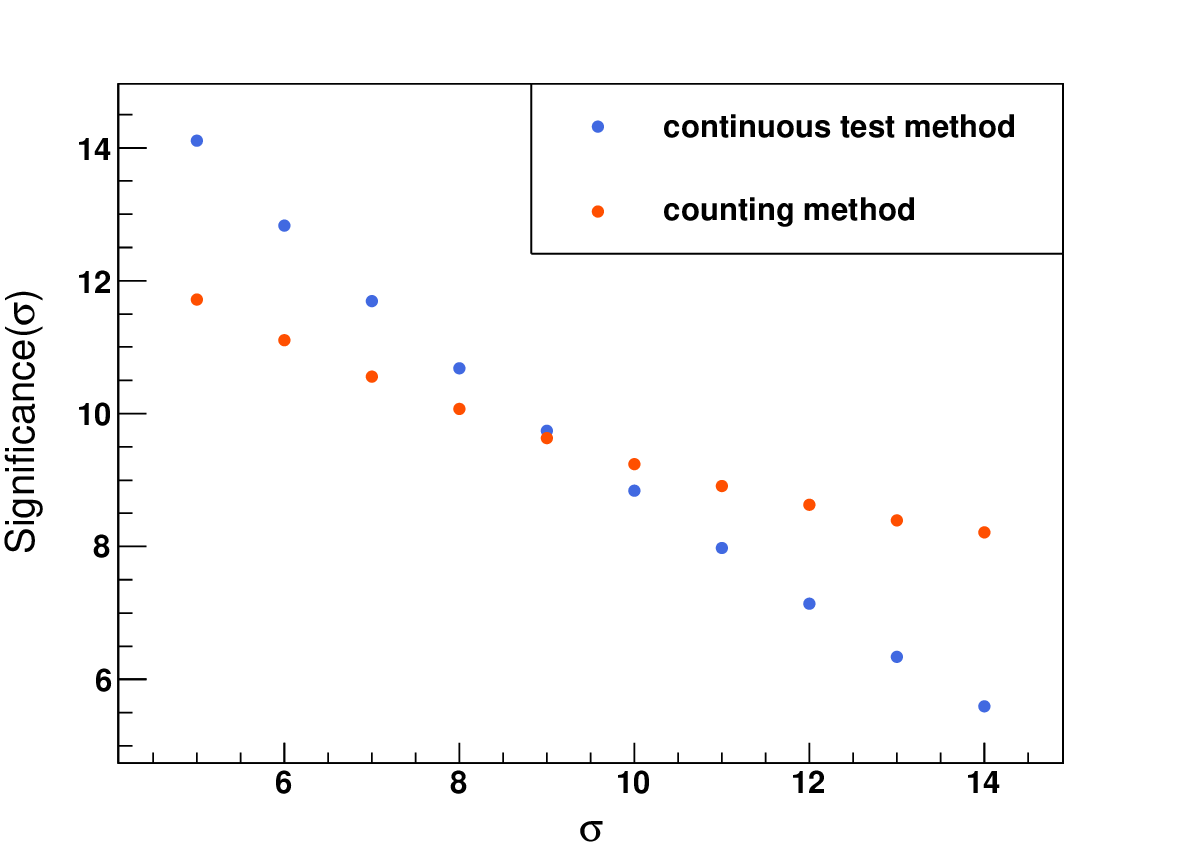}
\end{minipage}%
\caption{The significance distributions are calculated by counting method and continuous test method, respectively, with respect to different signal width, which are set with a sequence of varied values.}
\label{fig:compare}
\end{center}
\end{figure}

It would be intriguing to compare the effectiveness of the continuous test method and the counting method in estimating significance. Intuitively, the continuous test should offer advantages when the signal shape significantly differs from the background shape, as it provides additional information from the shapes. However, if the signal shape closely resembles that of the background, separating the signal and background from the fit would be challenging, making the counting method a better choice. To test this idea, we generated a series of samples with varying signal widths, i.e., $\sigma$ ranging from $5.0$ to $14.0$ in increments of $0.1$, and calculated the significance using the two methods described by Eqs.~\ref{eq:1} and~\ref{eq:2}. The results, depicted in Fig.~\ref{fig:compare}, confirm our expectations.

\section{Discussion}
After investigating the impact of various systematic uncertainties on significance, our conclusion is that these uncertainties typically lead to unpredictable effects on signal significance. Due to the complex correlation between cause and result, we recommend a conservative approach: estimate signal significance by conducting trials with all possible combinations of uncertainties associated with the fitting procedure, and then select the ``worst" outcome as the final result. Additionally, our study suggests that when the signal shape is distinguishable from the background, the continuous test method should be used for calculating signal significance, whereas the counting method is favored when the signal shape closes to the background. It's important to note that this study solely focuses on significance calculation for signals of known location, without considering the look-elsewhere effect, which is beyond the scope of this paper and may require specific extensive methods for further exploration.
\vspace{5cm}

% \section*{ACKNOWLEDGMENTS}

\end{document}